\documentclass[lettersize,journal]{IEEEtran}
\usepackage{amsmath,amsfonts}
\usepackage{algorithmic}
\usepackage{algorithm}
\usepackage{array}
\usepackage[caption=false,font=normalsize,labelfont=sf]{subfig}
\usepackage{textcomp}
\usepackage{stfloats}
\usepackage{url}
\usepackage{verbatim}
\usepackage{graphicx}
\usepackage{cite}

\usepackage{multirow}
\usepackage{array}

\usepackage{booktabs} % For professional looking tables
\usepackage{siunitx}  % For aligning numbers by decimal point
\usepackage{amssymb}

\newcommand{\etal}{\textit{et al}.}

\begin{document}

\title{Robust Multi-generation Learned Compression of \\ Point Cloud Attribute}

\author{Xiangzuo Liu\textsuperscript{*}, Zhikai Liu\textsuperscript{*}, Pengpeng Yu, Ruishan Huang, Fan Liang,~\IEEEmembership{Member,~IEEE}%
\thanks{Corresponding author: Fan Liang.}
\thanks{* Equal contribution.}
\thanks{Xiangzuo Liu, Zhikai Liu and Fan Liang are with the School of Electronics and Information Technology, Sun Yat-sen University, Guangzhou, Guangdong, 510006, China (e-mail: liuxz6@mail2.sysu.edu.cn; liuzhk6@mail2.sysu.edu.cn; isslf@sysu.edu.cn).}
\thanks{Pengpeng Yu is with the School of Electronics and Communication Engineering, the Shenzhen Campus of Sun Yat-sen University, Sun Yat-sen University, Shenzhen 518107, China (e-mail: yupp5@mail2.sysu.edu.cn).}
\thanks{Ruishan Huang is with the taixin-semi, Zhuhai, Guangdong, 519000, China (e-mail: huangruishan@taixin-semi.com).}
}

\markboth{Journal of \LaTeX\ Class Files,~Vol.~X, No.~X, Month~Year}%
{Shell \MakeLowercase{\textit{et al.}}: A Sample Article Using IEEEtran.cls for IEEE Journals}

\IEEEpubid{0000--0000/00\$00.00~\copyright~2021 IEEE}

\maketitle

\begin{abstract}
Existing learned point cloud attribute compression methods primarily focus on single-pass rate-distortion optimization, while overlooking the issue of cumulative distortion in multi-generation compression scenarios. This paper, for the first time, investigates the multi-generation issue in learned point cloud attribute compression. We identify two primary factors contributing to quality degradation in multi-generation compression: quantization-induced non-idempotency and transformation irreversibility. To address the former, we propose a Mapping Idempotency Constraint, that enables the network to learn the complete compression-decompression mapping, enhancing its robustness to repeated processes. To address the latter, we introduce a Transformation Reversibility Constraint, which preserves reversible information flow via a quantization-free training path. Further, we propose a Latent Variable Consistency Constraint which enhances the multi-generation compression robustness by incorporating a decompression-compression cross-generation path and a latent variable consistency loss term. Extensive experiments conducted on the Owlii and 8iVFB datasets verify that the proposed methods can effectively suppress multi-generation loss while maintaining single-pass rate-distortion performance comparable to baseline models.
\end{abstract}


\section{Introduction}
\IEEEPARstart{A}{s} a core representation of 3D scenes,
point clouds are indispensable in applications such as autonomous driving systems and immersive VR/AR experiences. 
Due to the voluminous and complex nature, point clouds pose fundamental challenges concerning the efficient representation, transmission and compression storage. The need for interoperability and effective data handling spurred initial research and standardization efforts in point cloud compression~\cite{ICRA12,Design17,JETCAS19MPEG,TCSVT21Ma}.

Point cloud compression faces a dual challenge: ensuring both the accurate reconstruction of geometric shapes and the fidelity of attribute information, such as color and reflectance, to maintain the visual and semantic integrity of scenes. Although MPEG standard methods (G-PCC~\cite{G-PCC} and V-PCC~\cite{V-PCC}) and deep learning models (e.g., Learned-PCGC~\cite{PCGC}, 3DAC~\cite{3DAC}, and FastPCC~\cite{FastPCC}) have achieved significant progress in single-pass compression, a critical yet often overlooked issue emerges in practical applications: distortion accumulation in iterative lossy compression cycles. Such multi-generation compression scenarios are common in iterative editing, cloud-based collaborative processing, cross-platform transmission and dynamic bitrate adaptation. In these scenarios, point clouds often undergo multiple encoding and decoding operations, leading to distortion accumulation.

\IEEEpubidadjcol

\begin{figure}[!t]
\centering
\includegraphics[width=3.5in]{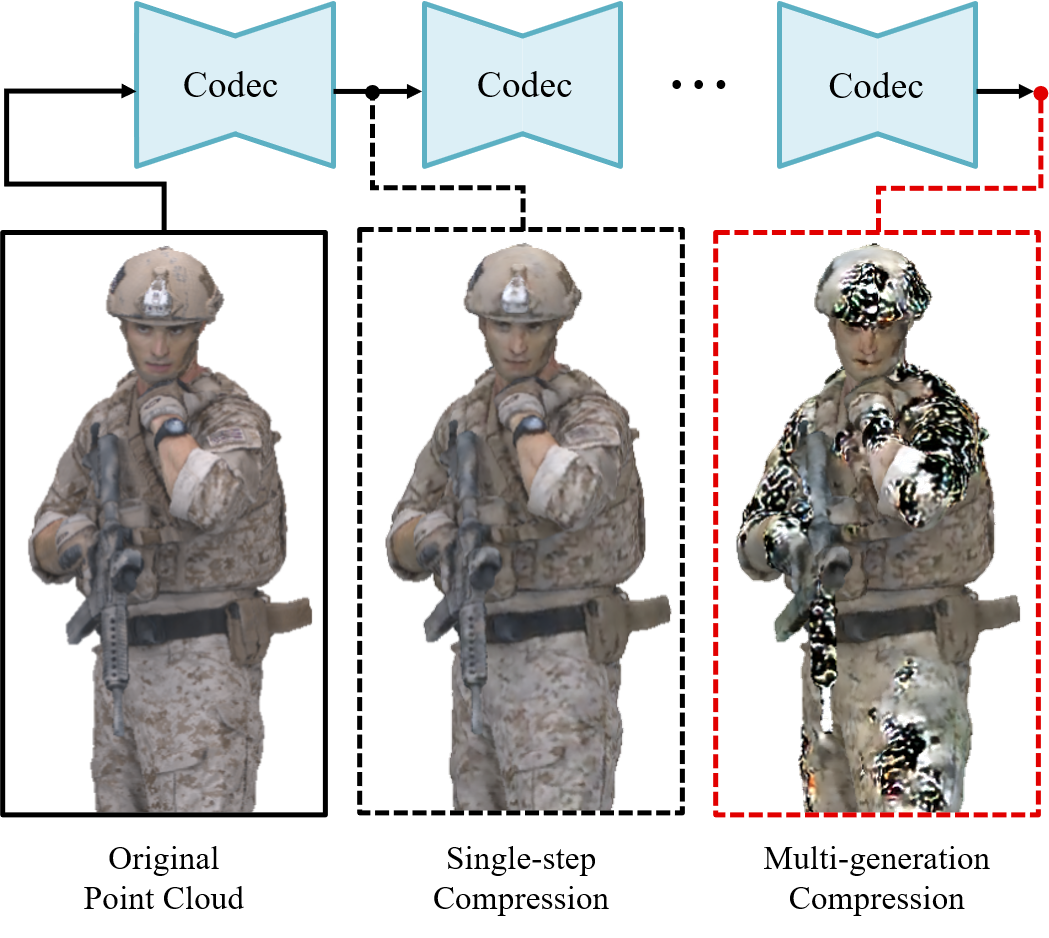}
\caption{Multi-generation compression results of learned point cloud attribute compression.}
\label{fig_1}
\end{figure}

As illustrated in Fig. \ref{fig_1}, learned point cloud compression frameworks suffer from quality degradation in multi-generation compression scenarios. This issue primarily stems from the synergistic effect of dynamic quantization error propagation and transform irreversibility. On one hand, the uncertainty introduced by the dynamic quantization mechanism, which is typically driven by a hyper-prior network, generates quantization errors. These errors accumulate and amplify with the number of iterations, causing inter-generational feature manifold drift. On the other hand, the symmetric structure of encoders and decoders loses its reversibility after the intervention of quantization operations, leading to a gradual distortion of the reconstructed space's manifold relative to the original space. The interaction of these two mechanisms results in irreversible quality degradation, including color distortion, detail loss and texture degradation, thereby severely impacting the applicability of point clouds in multi-generation compression scenarios.

To overcome the aforementioned limitations, this paper proposes a theoretical framework for analyzing multi-generation learned point cloud attribute compression. Furthermore, three constraint-based methods are proposed to optimize multi-generation robustness:

1) Mapping Idempotency Constraint (MIC): Reconstructs the training process to learn the complete mapping, including post-processing steps such as scaling and rounding, thereby enhancing the idempotency of the mapping in multi-generation compression.

2) Transformation Reversibility Constraint (TRC): Adds a quantization-free auxiliary path during training, leveraging the symmetry of the encoder-decoder structure to learn a reversible identity transformation, thus increasing the reversibility of the analysis/synthesis transforms.

3) Latent Variable Consistency Constraint (LCC): Introduces an inter-generational compression auxiliary path during training, inclusive of post-processing, to learn the mapping relationship of latent variables across generations, thereby enhancing the model's inter-generational decompression-compression capability.

\begin{figure}[!t]
\centering
\includegraphics[width=3.5in]{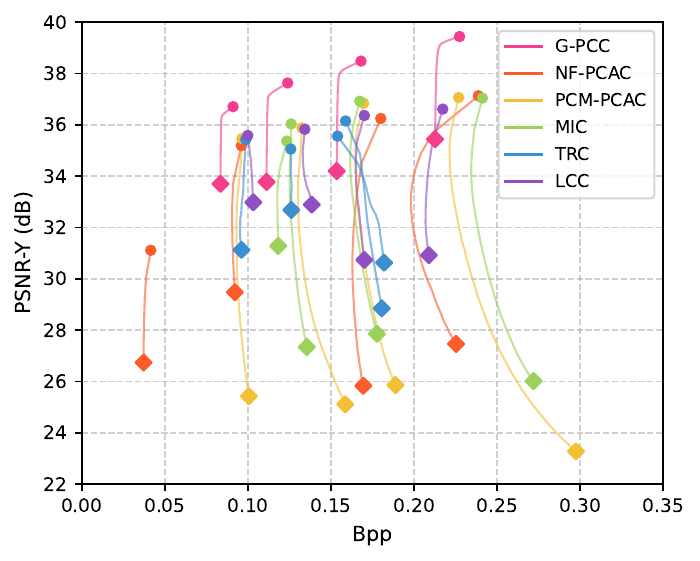}
\caption{Rate-distortion performance for different methods on the Owlii dataset. The dots ({\small{$\bullet$}}) and diamonds ({\small{$\Diamond$}}) represent the results of single-pass (1st) and 50-pass (50th) compression, respectively.}
\label{fig_2}
\end{figure}

Experiments show that our method can effectively enhance the multi-generation robustness of the model and is significantly better than previous work~\cite{NF-PCAC,PCM-PCAC}. Fig. \ref{fig_2} shows the excellent performance of the proposed methods. The remainder of this paper is organized as follows. Section~\ref{related} introduces related work on point cloud attribute compression and multi-generation robust image compression. Section~\ref{analysis} analyzes the mechanisms of multi-generation compression. Section~\ref{Method} describes our three proposed methods for optimizing multi-generation compression robustness. Section~\ref{experiment} provides a detailed overview of the experimental results and analysis. Finally, Section~\ref{conclusion} concludes the paper and discusses future work.

\section{Related Works}\label{related}
\subsection{Point Cloud Attribute Compression}
Point cloud compression techniques primarily address two types of data: geometric information, defined by 3D spatial coordinates, and attribute information, comprising visual features like color and reflectance. Mainstream traditional point cloud compression techniques can be categorized into two types: the G-PCC standard~\cite{G-PCC}, which directly encodes in 3D space, and the V-PCC standard~\cite{V-PCC}, based on 2D projection. The G-PCC standard includes three core attribute coding schemes: (1) Region Adaptive Hierarchical Transform (RAHT)~\cite{RAHT}, which achieves frequency-domain decorrelation through a geometry-aware hierarchical wavelet-like transform; (2) Predicting Transform~\cite{G-PCC}, which constructs a multi-level prediction structure based on level of detail and encodes the prediction residuals; and (3) Lifting Transform~\cite{lifting_trans}, implemented through alternating predict and update steps, where its update step improves the statistical properties of the signal by adjusting low-frequency components, thereby enhancing compression performance. The V-PCC standard employs a 3D-to-2D projection strategy, mapping point cloud attributes to 2D texture maps, %which are then compressed using HEVC/VVC video coding standards. 
which are then established video coding standards, such as H.265/HEVC~\cite{sullivan2012overview} or H.266/VVC~\cite{bross2021overview}. Although V-PCC can leverage mature video coding tools, the projection process can induce geometric distortions, leading to attribute aliasing, particularly noticeable in regions with complex surfaces~\cite{TMMVPCCArtifact}. It is noteworthy that while G-PCC maintains optimal attribute compression performance among current standards, its handcrafted transform models struggle to fully exploit the irregular distribution characteristics of point cloud data, limiting its development potential.

To overcome the limitations of traditional methods and further enhance compression performance, deep learning based methods have emerged. In recent years, learned point cloud attribute compression methods have achieved significant breakthroughs. Deep-PCAC, proposed by Sheng~\etal~\cite{Deep-PCAC}, is the first to construct an end-to-end compression framework, directly processing raw point clouds and avoiding voxelization or projection operations. Its core innovations include a second-order point convolution operator to capture long-range spatial correlations, a dense point initialization module to enhance feature propagation and a multi-scale loss function to optimize reconstruction quality. Sparse-PCAC, designed by Wang~\etal~\cite{Sparse-PCAC}, pioneers the application of sparse tensors for point cloud attribute representation. This method integrates sparse convolution with a hyper-prior entropy model, thereby enhancing entropy coding precision and achieving notable performance gains over the G-PCC reference software, TMC13 v6. NF-PCAC, proposed by Pinheiro~\etal~\cite{NF-PCAC}, innovatively employes a normalizing flow-based architecture. This approach uses reversible transformations to eliminate the low-dimensional bottleneck effect of variational autoencoders, significantly enhancing reconstruction fidelity at high bitrates. Huo~\etal~\cite{RO-PCAC} proposed RO-PCAC, a rendering-oriented compression framework, which integrates differentiable rendering with point cloud attribute compression and incorporates a Sparse Tensor Transformer (SP-Trans) to enhance feature analysis and synthesis capabilities, aiming to improve the quality of rendered images. PCM-PCAC, proposed in our previous work~\cite{PCM-PCAC}, designs a parallelized context model based on Morton sort. It utilizes cross-coordinate attention to model the correlation between geometric and attribute information, as well as global dependencies, simultaneously optimizing coding performance and decoding speed. %Guo~\etal~\cite{TSC-PCAC} introduced the TSC-PCAC method, which fuses Transformer and sparse convolution modules to achieve synergistic extraction of local dependencies and global features, culminating in state-of-the-art performance in several benchmark tests.

% Although the aforementioned methods have made significant progress in single-pass compression scenarios, none have considered the demand for successive encoding and decoding in practical applications. Consequently, in scenarios of multi-generation compression, cumulative distortions such as texture distortion and color shift will appear as shown in Fig. \ref{fig_1}, which exposes the theoretical defects of existing methods in multi-generation compression stability.
Although the aforementioned methods have made significant progress in single-pass compression scenarios, none have considered the demand for successive encoding and decoding in practical applications. Consequently, multi-generation compression inevitably leads to cumulative distortions, including texture degradation and color shifts, as illustrated in Fig. \ref{fig_1}, which exposes the theoretical defects of existing methods in multi-generation compression stability.

\subsection{Multi-Generation Robust Image Compression}
Traditional image encoders achieve multi-generation compression stability through reversible transforms and deterministic quantization. Sorial and Lynch~\cite{Sorial} systematically analyzed various factors contributing to multi-generation compression quality degradation, such as quantization, clipping and changes in coding parameters. JPEG~\cite{jpeg1991} and JPEG2000~\cite{jpeg20002001} employ scalar quantization mechanisms with fixed parameters, ensuring a deterministic distribution of variables within quantization intervals primarily through rounding operations, thereby satisfying idempotency requirements. Joshi~\etal~\cite{joshi} experimentally demonstrated that introducing the canvas system aligned with wavelet decomposition boundaries enables JPEG2000 to maintain superior idempotency in multi-generation compression. The JPEG XS standard~\cite{jpegXS2021} effectively suppresses multi-generation distortion caused by quantization and clamping operations through its head and toe region design~\cite{jpegXS17toeregion}. In the video coding domain, Erdem and Sezan~\cite{Erdem} proved the PSNR error saturation characteristic of the MPEG video compression algorithms in multi-generation compression using the theory of Generalized Projections, while Zhu~\etal~\cite{Zhu} achieved strict idempotency for H.264/AVC intra-frame coding through a clipping compensation matrix.

Learned image compression methods face new challenges regarding multi-generation robustness. Kim~\etal~\cite{Kim_c,Kim_j} were the first to systematically reveal the phenomenon of successive distortion accumulation in deep codecs and conducted detailed successive image compression benchmark tests. They proposed using a feature identity loss to constrain the inter-generational consistency of latent representations to improve the multi-generation compression stability of deep image compression, an idea that provided a reference for subsequent research on maintaining information stability within latent space. Li~\etal~\cite{Li_l} pointed out that quantization drift and transform irreversibility are crucial factors affecting the robustness of deep image compression. Based on this, they proposed a direct quantization strategy, a reversibility loss function, and a channel relaxation method to improve the multi-generation robustness of deep image compression. Among these, the focus on transform reversibility provided important insights for designing more multi-generation robust point cloud attribute codecs. Helminger~\etal~\cite{Helminger} proposed a deep image compression method based on a reversible normalizing flow architecture, establishing a bijective mapping between the latent space and the reconstructed image to achieve multi-generation compression idempotency. On a theoretical level, Li~\etal~\cite{Li_y} made a breakthrough in proving that reversibility is only a sufficient but not necessary condition for idempotence, and relaxed the bijective condition for achieving idempotence to right reversibility. Accordingly, they designed a right-inverse framework incorporating blocked convolution and null-space enhancement, achieving approximate idempotency in deep image compression. Zhang~\etal~\cite{Zhang_strg_idem} further extended the concept of ``strong idempotence'' to require codecs to maintain stability during multi-generation compression across different quality factors. These studies establish quantization determinism and transform reversibility as the theoretical foundations for multi-generation robustness.

In the field of point cloud compression, research specifically targeting multi-generation robust compression has not yet been conducted. Drawing inspiration from ideas in image compression, this paper establishes a comprehensive model for multi-generation point cloud attribute compression. It optimizes for the characteristics of point cloud data and the point cloud compression process, aiming to suppress the rapid decline in visual quality and attribute information distortion of point cloud attribute during multi-generation compression.

\section{Mechanism Analysis of Multi-generation Point Cloud Attribute Compression}\label{analysis}

\begin{figure*}[!t]
\centering
\includegraphics[width=7.16in]{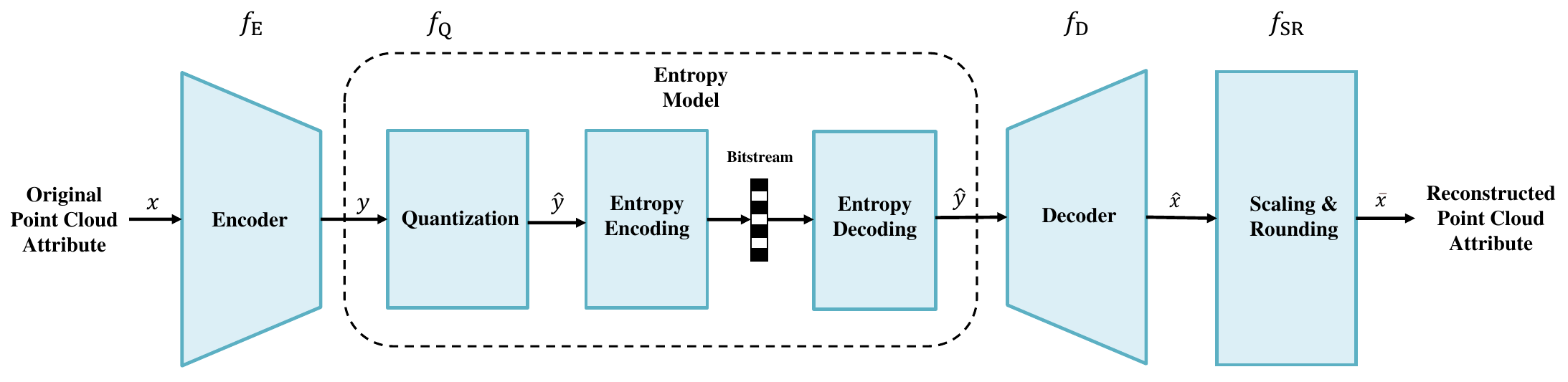}
\caption{A standard end-to-end framework for learned point cloud attribute compression, illustrating the main processing stages: an encoder ($f_\text{E}$), quantization ($f_\text{Q}$) and entropy encoding/decoding, a decoder ($f_\text{D}$), and a scaling \& rounding post-processing stage ($f_{\text{SR}}$).}
\label{fig_3}
\end{figure*}

This section delves into the underlying mechanisms of quality degradation in multi-generation point cloud attribute compression. First, the Section~\ref{section3A} formally defines the multi-generation compression process. Subsequently, Section~\ref{section3B} and Section~\ref{section3C} analyzes two primary factors that  cause non-idempotency in multi-generation compression: one is the quantization-induced non-idempotency and the other is the irreversibility in learned analysis and synthesis transformations. The synergistic effect of these two factors leads to distortion accumulation and quality degradation problems in multi-generation compression.

\subsection{Definition of Multi-generation Compression}\label{section3A}
Current learned point cloud attribute compression methods commonly adopt the end-to-end Variational Autoencoder (VAE) framework proposed by Ballé~\etal~\cite{Ballé} as their underlying architecture. This framework consists of five components: an encoder, a decoder, a hyper-encoder, a hyper-decoder, and an entropy model. In this framework, the hyper-encoder and hyper-decoder generate side information to provide conditional parameters for the entropy model. While this design effectively enhances entropy coding efficiency, its parameter generation process is independent of the main encoding-decoding path. Consequently, its direct impact on idempotency is considered secondary compared to the quantization and transformation stages in the main path, and thus it is simplified in our analysis.

As illustrated in Fig. \ref{fig_3}, mainstream learned point cloud attribute compression frameworks follow a processing chain of ``Encoding ($f_\text{E}$) - Quantization ($f_\text{Q}$) - Entropy encoding/decoding - Decoding ($f_\text{D}$) - Scaling and rounding ($f_{\text{SR}}$)''. The original point cloud $x$ is encoded into a latent variable $y$ through an analysis transform. This $y$ is then discretized into $\hat{y}$ via a centering quantization operation, which is subsequently entropy encoded into a compact bitstream. At the decoding end, the latent variable $\hat{y}$ is recovered through the reverse process, then decoded into reconstructed features $\hat{x}$ via a synthesis transform, and finally restored to the reconstructed point cloud $\bar{x}$ through post-processing operations. Therefore, we can represent the single-pass point cloud encoding-decoding process as:
\begin{equation}
\label{Eq1}
\bar{x} = f_{\text{SR}} \circ f_\text{D} \circ f_\text{Q} \circ f_\text{E} (x).
\end{equation}

The multi-generation point cloud compression process forms a Markov chain, where the states are the point cloud attributes. We use $f_{\text{SPCC}}(\cdot)$ to denote the single-pass compression operator in multi-generation point cloud compression. The $k$-th iterative process of multi-generation point cloud compression can then be modeled as:
\begin{equation}
\label{Eq2}
x_k = f_{\text{SPCC}}^k(x_0) = (f_{\text{SR}} \circ f_\text{D} \circ f_\text{Q} \circ f_\text{E})^k(x_0).
\end{equation}

In this study, we assume that the compression conditions are the same for each generation, meaning $f_{\text{SPCC}}(\cdot)$ remains constant. A key observation is that if $f_{\text{SPCC}}(\cdot)$ were an idempotent operator, meaning it satisfies the condition: 
\begin{equation}
\label{Eq2_1}
f_{\text{SPCC}}(\cdot) \circ f_{\text{SPCC}}(\cdot) = f_{\text{SPCC}}(\cdot), \end{equation} 
%then for any number of compression generations $k$ greater than one, 
then for any number of compression generations $k > 1$,
the output $x_k$ would be identical to $f_{\text{SPCC}}(x_0)$, the output of a initial single-pass compression. In other words, the reconstruction quality would remain stable across all subsequent generations, matching that of the first pass. However, as we will analyze, its constituent operations, particularly $f_\text{Q}$ and $f_{\text{SR}}$, prevent $f_{\text{SPCC}}(\cdot)$ from being an idempotent operator. This non-idempotency results in $x_k \neq x_1$ for $k > 1$, leading to a gradual degradation in the quality of multi-generation point cloud reconstruction.

\subsection{Quantization-induced Non-idempotency Issue}\label{section3B}
The idempotency of a compression model is inextricably linked to the quantization mechanism within the compression process. Traditional codecs like JPEG, for instance, achieve idempotency or approximate idempotency through fixed quantization tables and deterministic rounding, ensuring that once a value is quantized, it remains stable in subsequent identical compression cycles. However, current mainstream learned point cloud attribute compression models universally adopt the modified quantization strategy proposed by Minnen~\etal~\cite{Minnen} in their entropy models, as shown in the following equation:
\begin{equation}
\label{Eq3}
\hat{y} = f_\text{Q}(y) = \lfloor y - \mu \rceil + \mu,
\end{equation}
where $\lfloor \cdot \rceil$ represents the rounding operation, and $\mu$ is typically generated dynamically by a hyper-prior network based on the current input. It is noteworthy that if $\mu$ is considered given, meaning it is fixed in the $f_\text{Q}$ operation, this quantization operator $f_\text{Q}$ itself is idempotent, i.e., $f_\text{Q}(f_\text{Q}(y)) = f_\text{Q}(y)$. However, the critical issue is that $f_\text{Q}$ is generally a lossy operation: for most continuous-valued latent variables $y$ and hyper-prior estimates $\mu$, because $y - \mu \notin \mathbb{Z}$, where $\mathbb{Z}$ represents the set of integers, the quantized $\hat{y}$ is not equal to the original $y$, i.e., $f_\text{Q}(y) \neq y$.

Due to the lossy nature of $f_\text{Q}$, the output $x_1 = f_{\text{SPCC}}(x_0)$ obtained after one complete encoding-decoding pass is typically not equal to the original input $x_0$. When this altered $x_1$ is used as the input for the next compression generation, its corresponding latent variable $y_1 = f_\text{E}(x_1)$ will deviate from $y_0 = f_\text{E}(x_0)$, consequently causing subsequent quantization results and reconstruction outcomes to change as well. This cumulative effect results in the entire compression operator $f_{\text{SPCC}}(\cdot)$ not being idempotent:
\begin{equation}
\label{Eq4}
\exists\ x_0 \text{ s.t. } f_{\text{SPCC}}(f_{\text{SPCC}}(x_0)) \neq f_{\text{SPCC}}(x_0).
\end{equation}

This non-idempotency of the overall compression process manifests specifically as a distortion accumulation phenomenon in multi-generation compression: During the $k$-th generation compression, the input to the encoder, $x_{k-1}$, has deviated compared to the input of the previous generation, $x_{k-2}$. This leads to changes in its corresponding latent variable $y_{k-1} = f_\text{E}(x_{k-1})$ and the mean estimate $\mu_{k-1}$ generated by the hyper-prior network. These factors collectively contribute to the gradual accumulation and exacerbation of distortion.

In addition to the quantization process within the entropy model, the post-processing of point cloud attributes also introduces attribute-specific information loss. The diversity of point cloud attributes necessitates customized handling for different data types during post-processing, which is distinctly different from image/video compression. Point cloud attributes such as color, normal vectors, and reflectance intensity undergo post-processing to restore their data type and magnitude during reconstruction. Examples include scaling and rounding operations for color information, mapping operations for normal vectors, and bit-depth limiting operations for reflectance intensity. While these attribute-specific post-processing operations might only introduce minor errors in single-pass compression, their cumulative effect can significantly reduce system stability and increase the model's non-idempotency in multi-generation compression scenarios if the errors are not strictly confined within quantization intervals.

This paper focuses on color attributes as the primary research subject due to their combined characteristics of: 1) sensitivity to the data type conversion from uint8 to float32 and then back to uint8; 2) non-linear error amplification properties; and 3) high perceptibility by the human visual system. The scaling and rounding operations in the color attributes encoding-decoding process can be represented as:
\begin{equation}
\label{Eq5}
f_{\text{SR}}(x) = \lfloor \text{clip}(x) \times 255 \rceil,
\end{equation}
where $\text{clip}(\cdot)$ denotes an operation that clips all values of $x$ to ensure they fall within the range [0, 1], and $\lfloor \cdot \rceil$ represents the rounding operation. In single-pass compression, if the reconstruction error $\delta$ satisfies the following condition:
\begin{equation}
\label{Eq5_0}
\delta= \left| \widehat{x}- \frac{x_{0}}{255} \right| \leq \frac{0.5}{255},
\end{equation}
the rounding operation can ensure $f_{\text{SR}}(\hat{x}) = x_0$. However, in multi-generation compression scenarios, accumulated errors from previous generations can cause the input $x$ to exceed the current generation's quantization interval, triggering an error step phenomenon. For example, if the error in the $k$-th generation is $\delta_{k} = 0.6/255$, the rounded result will deviate from the true value by one unit. This non-linear distortion can form a resonance effect through iterative compression, leading to a rapid decline in the reconstruction quality of point cloud attributes.

\subsection{Irreversibility Issue of Learned Transformation}\label{section3C}
Research by Li~\etal~\cite{Li_l} has analyzed how the reversibility of the transformation in image compression directly impacts quality loss in multi-generation compression. Furthermore, Li~\etal~\cite{Li_y} relaxed the sufficient condition for codecs to achieve idempotency from reversibility to right-invertibility. In point cloud attribute compression, performance is similarly affected. Compared to learned methods, traditional methods like G-PCC exhibit excellent multi-generation stability, as shown in Fig. \ref{fig_2}, which is partly attributable to its use of handcrafted reversible transformations.

Learned methods perceives and construct analysis and synthesis transformations via neural networks. During training, this process can be approximately represented by the following equation:
\begin{equation}
\label{Eq6}
\hat{x} = f_\text{D} \circ f_\text{Q} \circ f_\text{E} (x).
\end{equation}

In training, the compression model continuously trains the parameters of the transformation networks within $f_\text{E}$ and $f_\text{D}$ by constraining the MSE between $x$ and $ \hat{x}$. The quantization process $f_\text{Q}$ is often simulated by adding noise. The essence of $f_\text{Q}$ is an irreversible discretization of continuous features; when the information flow passes through $f_\text{Q}$, it inevitably leads to irreversible information loss in the feature space. $f_\text{E}$ and $f_D$ typically adopt symmetric structures. Assuming training under ideal conditions without quantization interference, that is, $f_\text{Q}$ is removed, the composite mapping $f_\text{E}$ and $f_\text{D}$ approximates an identity transformation $x = f_\text{D} \circ f_\text{E} (x)$. This would satisfy the reversibility condition $f_\text{D} = f_\text{E}^{-1}$. However, during actual deployment, the quantization operation $f_\text{Q}$ projects the continuous latent space manifold $\mathcal{M}_Y$ onto a discrete set of Lattice points $\mathbb{L}$:
\begin{equation}
\label{Eq6_1}
\mathbb{L} = \{\hat{y} \ |\ \hat{y} = \lfloor y - \mu \rceil + \mu \}.
\end{equation}

Consequently, the input manifold to the decoder becomes $\mathcal{M}_Q = f_\text{Q}(\mathcal{M}_Y)$. The decoder $f_\text{D}$ then needs to learn a non-surjective mapping from $\mathcal{M}_Q$ to the original manifold $\mathcal{M}_0$, rather than recovering the strict inverse transform of $f_\text{E}$. This irreversibility leads to two levels of problems: 1) The Hausdorff distance between the reconstructed manifold $\mathcal{M}_r = f_\text{D}(\mathcal{M}_Q)$ and the original manifold $\mathcal{M}_0$ increases with multi-generation compression generations; 2) The topology of the latent space undergoes irreversible distortion, weakening the feature representation capability and its utility for accurate reconstruction. The coupled effect of these issues results in a continuous decline in the quality of point cloud attributes in multi-generation compression scenarios.

\section{Methodology}\label{Method}

\begin{figure*}[!t]
\centering
\includegraphics[width=7.16in]{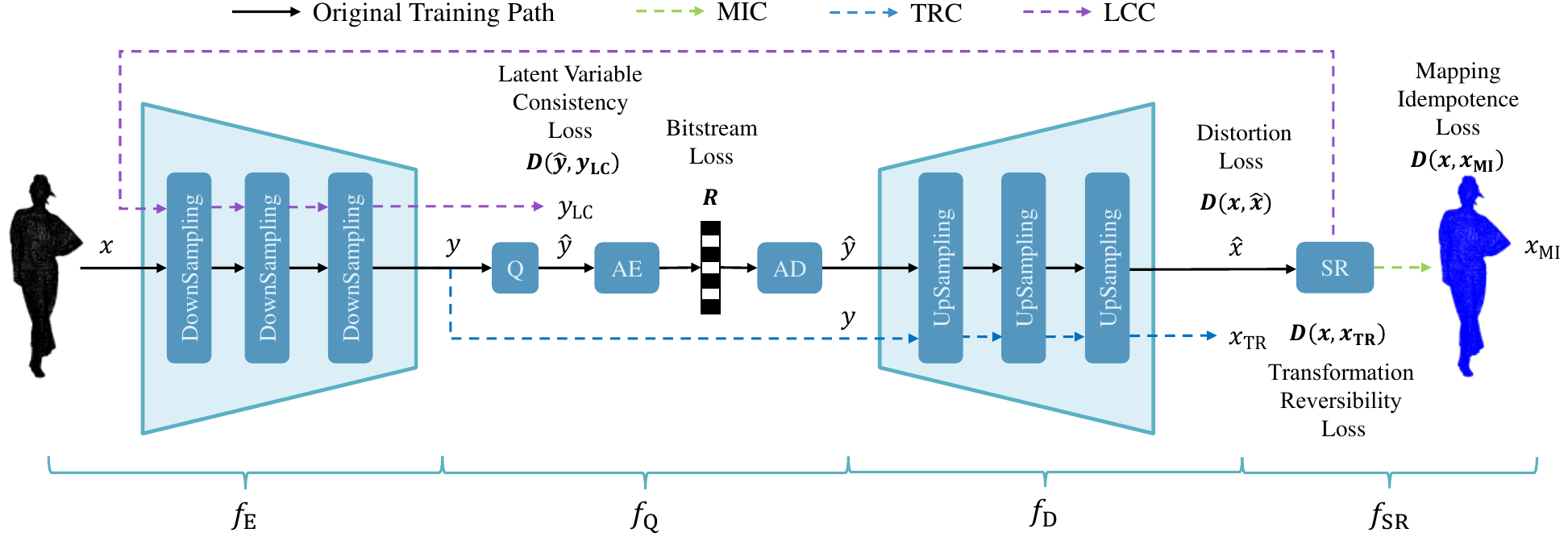}
\caption{Overview of the proposed training framework for enhancing multi-generation compression robustness, incorporating MIC (green dashed path), TRC (blue dashed path), and LCC (purple dashed path).}
\label{fig_4}
\end{figure*}

Fig. \ref{fig_4} illustrates the single-pass training process of our proposed method for point cloud attribute compression, which incorporates three methods to enhance multi-generation compression robustness: the Mapping Idempotency Constraint (MIC), the Transformation Reversibility Constraint (TRC), and the Latent Variable Consistency Constraint (LCC).

\subsection{Mapping Idempotence Constraint}
In Section~\ref{section3B}, we discussed how quantization-involved operations in the post-processing of point cloud attribute compression can inhibit multi-generation compression robustness. In the training process of mainstream point cloud compression models, the distortion loss is calculated only on the reconstructed values of type float32 within a normalized space, while scaling and rounding operations are typically treated as post-processing modules independent of the neural network. While this decoupled training method can simplify the backpropagation process, it leads to a lack of completeness in the encoder-decoder mapping relationship: the model fails to learn the end-to-end mapping from the original format to the reconstructed storage format, which entails the data type conversion from uint8 to float32 and then back to uint8. 

To address this core issue, this paper proposes a Mapping Idempotency Constraint (MIC), which reconstructs the training process to achieve an operational closed loop. Specifically, the model executes the following training process to replace the training process in \eqref{Eq6}:
\begin{equation}
\label{Eq7}
x_{\text{MI}} = f_{\text{SR}} \circ f_\text{D} \circ f_\text{Q} \circ f_\text{E} (x).
\end{equation}

This process strictly simulates the complete mapping path during actual deployment, converting the reconstructed value $\hat{x}$ back to the original space of uint8 type to generate $x_{\text{MI}}$. As shown in Fig. \ref{fig_4}, during this process, we design a mapping idempotency loss term: 
\begin{equation}
\label{Eq8}
\mathcal{L}_{\text{MI}} = D(x, x_{\text{MI}}) = ||x - x_{\text{MI}}||_2.
\end{equation}

And we use $\mathcal{L}_{\text{MI}}$ to replace the distortion loss $\mathcal{L}_D$ in the original loss function. Its effect is to guide the model to learn the complete compression-decompression mapping process, improving the idempotency of the mapping in multi-generation compression. The complete loss function in this method can be expressed as: 
\begin{equation}
\label{Eq9}
\mathcal{L} = \mathcal{L}_{Rate} + \lambda \mathcal{L}_{\text{MI}},
\end{equation}
where $\mathcal{L}_{Rate}$ represents the rate loss term calculated by the cross-entropy loss function, and $\lambda$ is a hyperparameter that balances the loss terms. During training, to overcome the gradient discontinuity problem caused by quantization steps, we introduce a Straight-Through Estimator (STE)~\cite{STE}. This sets the gradient of the rounding operation to 1 during backpropagation, allowing the neural network to bypass the non-differentiability of the rounding operation during backpropagation and ensuring that gradients can propagate through the entire closed-loop path to update network parameters.

\subsection{Transformation Reversibility Constraint}
In Section~\ref{section3C}, we concluded that the irreversibility of transformation in point cloud attribute compression exacerbates the instability of multi-generation point cloud compression. The encoder and decoder of existing learned compression models are typically designed with symmetric structures. However, the intervention of quantization operations disrupts this symmetry, leading to irreversible distortion of the latent features. Inspired by the reversibility loss function proposed by Li~\etal~\cite{Li_l} in the field of image compression, this paper designs a Transformation Reversibility Constraint (TRC) tailored to the characteristics of point cloud attributes. This constraint aims to compel the network to leverage its inherent symmetric structure to learn a reversible analysis-synthesis transformation pair, under conditions free from the interference of the quantization operation $f_\text{Q}$.

To this end, we propose a dual-path training approach, introducing a transformation reversibility loss term $\mathcal{L}_{\text{TR}}$ during the training process. As shown in Fig. \ref{fig_4}, while the main path, containing the quantization operation, continues to perform rate-distortion (RD) optimization, a quantization-free auxiliary path is constructed:
\begin{equation}
\label{Eq10}
x_{\text{TR}} = f_\text{D} \circ f_\text{E} (x).
\end{equation}

This path completely bypasses the quantization and entropy coding modules, directly learning an identity mapping $x = f_\text{D} \circ f_\text{E} (x)$ that is undisturbed by quantization. The deviation between the attributes of the quantization-free reconstructed point cloud and those of the original point cloud is penalized using mean squared error. The transformation reversibility loss term can be expressed as:
\begin{equation}
\label{Eq11}
\mathcal{L}_{\text{TR}} = D(x, x_{\text{TR}}) = ||x - x_{\text{TR}}||_2.
\end{equation}

By compelling the codec to achieve exact reconstruction under quantization-free conditions, the network is motivated to learn the underlying mapping symmetry. After incorporating the TRC method, the loss function becomes:
\begin{equation}
\label{Eq12}
\mathcal{L} = \mathcal{L}_{Rate} + \lambda \mathcal{L}_D + \alpha \mathcal{L}_{\text{TR}},
\end{equation}
where $\lambda$ and $\alpha$ are hyperparameters that balance the loss terms.

\subsection{Latent Variable Consistency Constraint}
The essence of multi-generation compression is a dynamic iterative system, where minor deviations in the latent representation can be non-linearly amplified over multiple compression cycles. The current mainstream methods solely focus on single-pass RD performance, neglecting the cross-generational evolution of latent variables. Inspired by the feature identity loss method of Kim~\etal~\cite{Kim_c,Kim_j}, we propose a Latent Variable Consistency Constraint (LCC) to suppress inter-generational error propagation. This constraint ensures that latent variables remain consistent after undergoing an inter-generational decompression-compression process. We construct an inter-generational compression auxiliary path:
\begin{equation}
\label{Eq13}
\hat{y}_{\text{LC}} = f_\text{E} \circ f_{\text{SR}} \circ f_\text{D} (\hat{y}).
\end{equation}

This constraint enables the network to learn the mapping process that includes post-processing operations, while also perceiving the inter-generational mapping process of decompression-compression. As illustrated in Fig. \ref{fig_4}, the inter-generational compression process from (\ref{Eq13}) is added during training, and a latent variable consistency loss $\mathcal{L}_{\text{LC}}$ is established to enforce the latent variable consistency constraint. To measure the consistency of the latent space across generations, we employ mean squared error as the metric and design the latent variable consistency loss as:
\begin{equation}
\label{Eq14}
\mathcal{L}_{\text{LC}} = D(\hat{y}, \hat{y}_{\text{LC}}) = ||\hat{y} - \hat{y}_{\text{LC}}||_2.
\end{equation}

This constraint suppresses the resonance effect caused by step errors by enforcing consistency of latent variables between adjacent generations. Unlike the feature identity loss method, we incorporate post-processing operations tailored to the characteristics of point cloud attributes. Furthermore, the latent variable consistency constraint does not include the quantization process $f_\text{Q}$ of the entropy model, aiming to enhance the reversibility of the transformations. Finally, the loss function incorporating the LCC method is formulated as follows:
\begin{equation}
\label{Eq15}
\mathcal{L} = \mathcal{L}_{Rate} + \lambda \mathcal{L}_D + \beta \mathcal{L}_{\text{LC}},
\end{equation}
where $\lambda$ and $\beta$ are hyperparameters used to balance the weights of the respective loss terms. Similar to the Mapping Idempotency Constraint, we add a STE during the training process of the Latent Variable Consistency Constraint to ensure the differentiability of the rounding operation during backpropagation.

\section{Experiments}\label{experiment}
In this section, we conduct qualitative and quantitative tests on standard datasets to validate the effectiveness of the proposed methods. Furthermore, we analyze the independence of the proposed methods and the feasibility of their combination through ablation studies.

\subsection{Experiment Setup}
\subsubsection{Compared Methods}
This study evaluates the performance of three representative methods in multi-generation point cloud attribute compression scenarios: the traditional standard G-PCC (TMC13 v23, RAHT)~\cite{G-PCCv23}, the normalizing flow-based NF-PCAC~\cite{NF-PCAC}, and the VAE framework-based PCM-PCAC~\cite{PCM-PCAC}. Given that our previous work, PCM-PCAC, employs a representative VAE architecture, it is selected as the baseline model for in-depth research and improvement of generational distortion issues in point cloud attribute compression. The proposed Mapping Idempotence Constraint (MIC), Transformation Reversibility Constraint (TRC), and Latent Variable Consistency Constraint (LCC) are all implemented and evaluated on the PCM-PCAC framework. The effectiveness of the proposed methods is validated through a comprehensive comparison with these three anchor methods.

\subsubsection{Benchmark Datasets}
The datasets used in this experiment include 8i Voxelized Full Bodies (8iVFB)~\cite{8i}, Owlii~\cite{Owlii} and Real World Textured Things (RWTT)~\cite{rwtt}. 8iVFB contains four dynamic voxelized full body point cloud sequences with a resolution of 1024. Owlii comprises four dynamic human textured mesh sequences with a resolution of 2048. RWTT includes 568 textured 3D models generated by photo reconstruction tools. In this experiment, the reference software mpeg-pcc-mmetric~\cite{mmetric} is used to convert mesh models in obj format to a unified ply format. 8iVFB and Owlii are used as test datasets, with specific test point cloud samples detailed in Table~\ref{tab1}. RWTT is used as the training dataset, with training performed on 404 RWTT objects% selected in MPEG's CTC on AI-based point cloud coding~\cite{CTC}.
, selected according to MPEG's Common Test Conditions (CTC) for AI-based point cloud coding~\cite{CTC}.

% \begin{table}
% \begin{center}
% \caption{Point Cloud Samples Used for Testing}
% \label{tab1}
% \begin{tabular}{| c | c | c |}
% \cline{1-1}
% \hline
% \multicolumn{1}{|c|}{Sample name}   & Resolution & Number of points \\ \cline{1-1}
% \hline
% basketball\_player\_vox11\_00000001 & 2048       & 2880057          \\
% \hline
% dancer\_vox11\_00000001             & 2048       & 2592758          \\
% \hline
% exercise\_vox11\_00000001           & 2048       & 2391718          \\
% \hline
% model\_vox11\_00000001              & 2048       & 2458429          \\
% \hline
% longdress\_vox10\_1051              & 1024       & 765821           \\
% \hline
% loot\_vox10\_1000                   & 1024       & 784142           \\
% \hline
% redandblack\_vox10\_1450            & 1024       & 729133           \\
% \hline
% soldier\_vox10\_0536                & 1024       & 1059810          \\
% \hline
% \end{tabular}
% \end{center}
% \end{table}
\begin{table}[tb] % 您可以选择合适的浮动位置参数
\centering % 使表格居中
\captionsetup{skip=5pt} % 如果需要，可以取消注释以增加标题和表格间距 (需要 caption 宏包)
\caption{Point Cloud Samples Used for Testing}
\label{tab1} % 您的标签
\renewcommand{\arraystretch}{1.2} % 可选: 略微增加行高，使表格更疏朗
% l: 左对齐 (Sample name)
% S[table-format=4.0]: 数字列，整数最多4位 (Resolution)
% S[table-format=7.0]: 数字列，整数最多7位 (Number of points)
% @{} 用于去除列两边的额外空白
\begin{tabular}{@{}l S[table-format=4.0] S[table-format=7.0]@{}}
\toprule
Sample name                        & {Resolution} & {Number of points} \\ % S列的表头文本需要用{}括起来
\midrule
basketball\_player\_vox11\_00000001 & 2048         & 2880057            \\
dancer\_vox11\_00000001             & 2048         & 2592758            \\
exercise\_vox11\_00000001           & 2048         & 2391718            \\
model\_vox11\_00000001              & 2048         & 2458429            \\
longdress\_vox10\_1051              & 1024         & 765821             \\
loot\_vox10\_1000                   & 1024         & 784142             \\
redandblack\_vox10\_1450            & 1024         & 729133             \\
soldier\_vox10\_0536                & 1024         & 1059810            \\
\bottomrule
\end{tabular}
% \end{center} % \centering 命令已经使其居中，不再需要 center 环境
\end{table}

\subsubsection{Evaluation Metrics}
This study employs the following metrics to evaluate the performance of point cloud attribute compression. Primarily, the Peak Signal-to-Noise Ratio of the Y component (PSNR-Y) of color attributes, calculated using the reference software mpeg-pcc-dmetric~\cite{dmetric}, is used as the core metric for the quality of reconstructed point clouds. Bits per point (Bpp) is used to measure the size of the compressed bitstream. Addressing the specificity of multi-generation compression scenarios, and drawing upon the work of Li~\etal~\cite{Li_y}, we use the PSNR-Y Drop to assess the multi-generation compression robustness of different methods. After multiple successive compression cycles, a smaller PSNR-Y Drop value indicates that the method possesses superior multi-generation compression robustness.

Furthermore, to more comprehensively evaluate the characteristics of compression methods in multi-generation scenarios, this paper introduces two novel metrics: $\Delta$PSNR-Y and Drop Convergence Rate. 

$\Delta$PSNR-Y represents the change in PSNR-Y of the reconstructed point cloud attributes in the current compression generation relative to the previous generation. It can intuitively reflect the instantaneous stability and quality fluctuation of a model at various stages of the multi-generation compression process. Specifically, for the $k$-th generation, it is defined as:
\begin{equation}
\label{eq:delta_psnr_y}
\Delta\text{PSNR-Y}_{k} = \text{PSNR-Y}_{k-1} - \text{PSNR-Y}_{k},
\end{equation}
where $\text{PSNR-Y}_{k}$ represents the PSNR-Y value of the reconstructed attributes after the $k$-th compression generation.
 
The Drop Convergence Rate is designed to more clearly illustrate the convergence characteristics of compression methods in multi-generation scenarios, and its calculation formula is as follows: 
\begin{equation}
\label{Eq16}
\text{Drop Convergence Rate} = \ln\left(\frac{\Delta \text{PSNR-Y}}{\text{Max PSNR-Y Drop}}\right),
\end{equation}
where the normalization factor, Max PSNR-Y Drop, is the largest PSNR-Y Drop observed among all compared methods after a specific number of compression iterations. The purpose of the logarithmic transformation is to amplify the differences in convergence speeds among various methods, thereby more significantly distinguishing their long-term stability. A smaller Drop Convergence Rate typically signifies faster and more stable convergence performance.

\subsubsection{Training Settings}
To alleviate memory pressure during large-scale point cloud training and to accelerate training speed, we implement dynamic random cropping on the original point cloud data during training. The specific cropping strategy is: based on KD-Tree recursive spatial partitioning, the dimension with the largest variance is selected at each iteration. A random threshold interval is then applied to retain half of point clouds, which is equivalent to a scaling factor of $0.5$. This process is executed recursively until a sub-point cloud block with fewer than K points is obtained. This sub-point cloud block, which preserves complete local structures, serves as the training data for the model. In this experiment, K is set to 250,000.

In this experiment, $\lambda$ is set to 6000, 4000, 2000, and 1000. These values are chosen to cover a bitrate range corresponding to the test bitrate points of the anchor methods, facilitating a fair performance comparison. The batch size is set to 4. The Adam optimizer is used for training, with the learning rate starting at $1 \times 10^{-4}$ and decreasing by 20\% every 20 epochs. Each model is trained for 200 epochs. The experiments were conducted on a server equipped with an Intel Xeon Gold 6238 CPU and an NVIDIA TITAN RTX GPU.

\subsection{Overall Evaluation}
In the experiment, we implemented MIC, TRC and LCC respectively. In the experiments of TRC and LCC, we empirically set $\alpha=\lambda$, $\beta=100$ for training respectively. Such setting can maintain the single-pass compression rate distortion performance similar to the original method. We compared the performance of the proposed method with PCM-PCAC, NF-PCAC and G-PCC on the Owlii dataset and 8iVFB dataset for 50-pass successive compressions.

\subsubsection{Qualitative Comparisons}
\begin{figure*}[!t]
	\centering
		\centering
          \subfloat[Owlii: dancer\_vox11\_00000001\label{fig_5a}]{
        \includegraphics[width=7.16in]{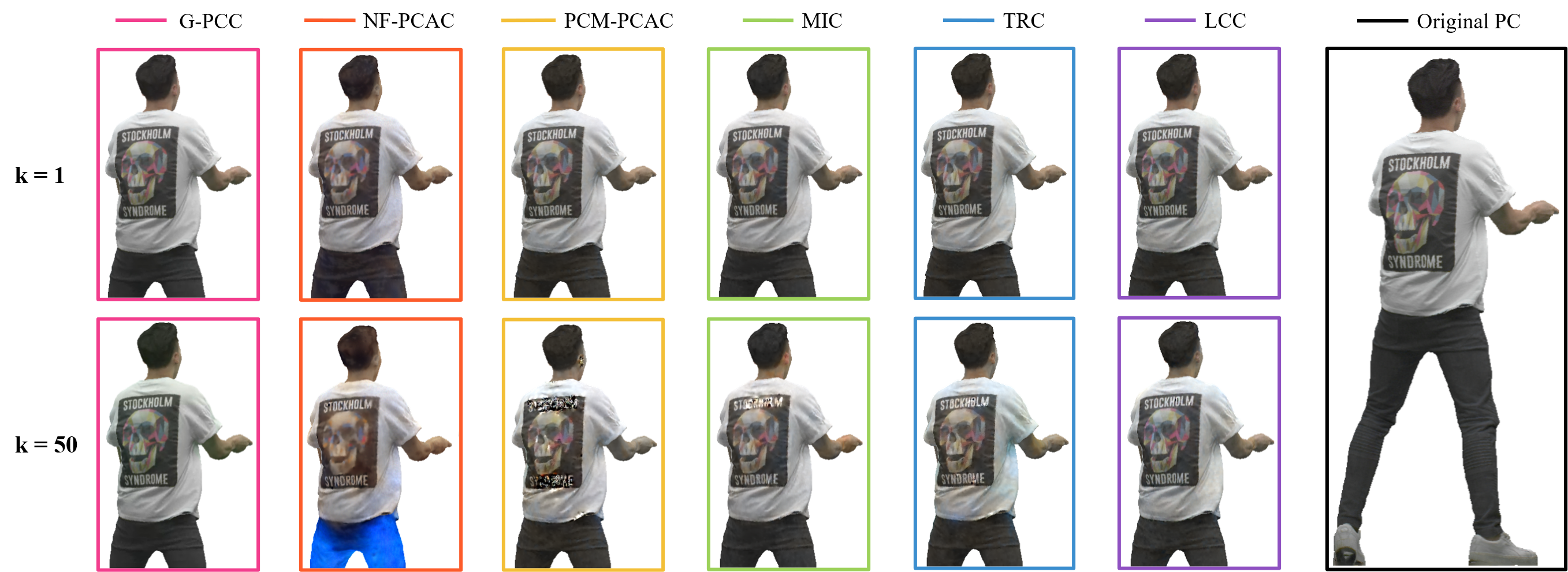}
      }
		% \includegraphics[width=\textwidth]{Fig/Fig5_1.png}
		% \subfloat{Owlii: dancer\_vox11\_00000001}		
\\
        \centering
              \subfloat[8iVFB: loot\_vox10\_1000\label{fig_5b}]{
        \includegraphics[width=7.16in]{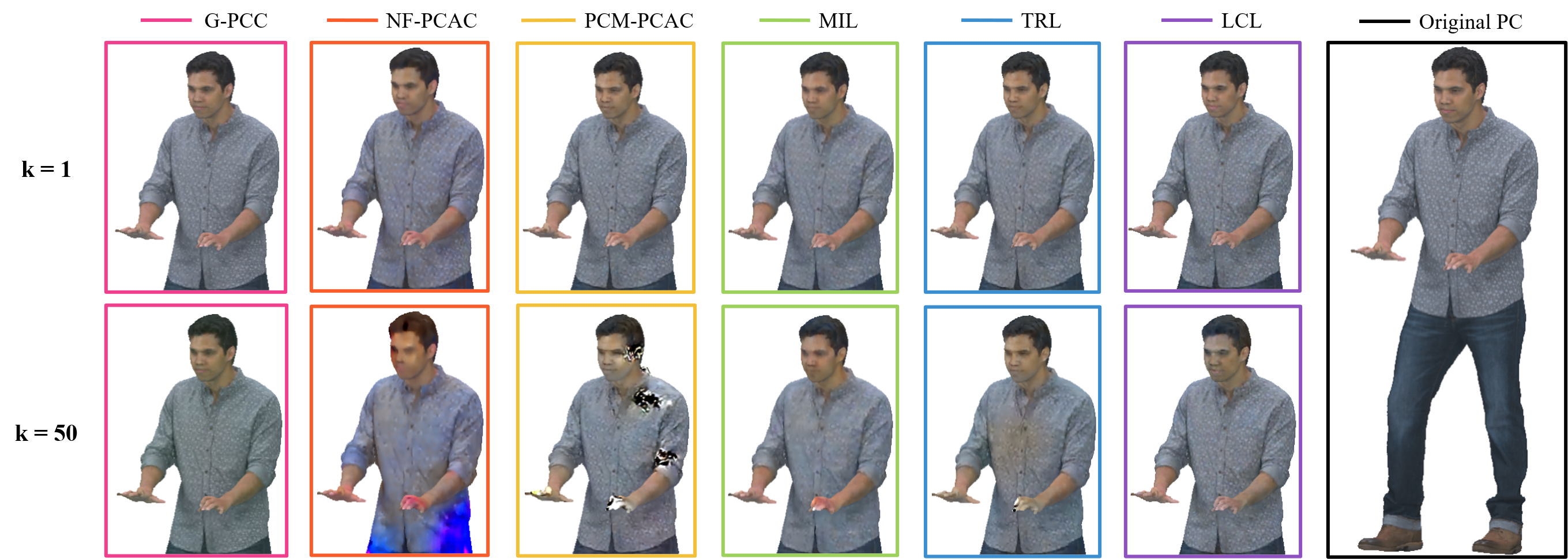}
      }
    		% \includegraphics[width=\textwidth]{Fig/Fig5_1.png}
    		% \subfloat{Owlii: dancer\_vox11\_00000001}
    \caption{Visual comparison of single-pass compression and 50-pass successive compression results of different methods on Owlii and 8iVFB.}
    \label{fig_5}
\end{figure*}

Maintaining visual quality is a crucial task in point cloud attribute compression. To comprehensively evaluate the performance of various point cloud attribute compression methods in multi-generation compression scenarios, we selected a point cloud sample from each test dataset. For each of the six methods, we chose a set of models with similar single-pass RD performance for multi-generation compression and visualized the reconstruction results. Fig. \ref{fig_5} shows the visualized images of point cloud samples after single-pass compression and 50-pass successive compression.

% \begin{figure*}[!t]
%     \centering
%     \subfloat[Owlii: dancer\_vox11\_00000001\label{fig_5a}]
%     {%
%         \includegraphics[width=7.16in]{Fig/Fig5_1.png}
%     }\\
%     \subfloat[8iVFB: loot\_vox10\_1000\label{fig_5b}]
%     {%
%     \includegraphics[width=7.16in]{Fig/Fig5_2.png}
%     }
%     \caption{Visual comparison of single-pass compression and 50-pass successive compression results of different methods on Owlii and 8iVFB.\label{fig_5}}
% \end{figure*}

Based on visual effect comparisons, we can observe the visual degradation characteristics of the six methods. G-PCC maintains good overall brightness and texture details after 50 compression rounds, with only slight color shifts, demonstrating the good multi-generation robustness of traditional point cloud attribute compression methods. Both learned methods, NF-PCAC and PCM-PCAC, exhibit significant distortion. Specifically, NF-PCAC shows local blue artifacts even after the initial compression, a phenomenon that intensifies with increasing iterations, resulting in severe color distortion after 50 compression rounds. PCM-PCAC has bad blocks after 50 compression rounds, particularly in areas with complex textures, such as the text outline shown in Fig. \ref{fig_5a} and the cloth folds depicted in Fig. \ref{fig_5b}, where this phenomenon is more pronounced.

Among the proposed methods, MIC alleviates most of the blocking artifacts that appear in texture-rich areas, but it loses detail information after 50 compression rounds, resulting in local blurring. TRC further suppresses regional pixel damage but introduces local color deviations, such as the blue artifacts visible in Fig. \ref{fig_5a} and the red artifacts apparent in Fig. \ref{fig_5b}. In contrast, LCC achieves a satisfactory improvement in multi-generation robustness, maintaining visual quality close to that of single-pass compression after 50 rounds. Its texture fidelity and color fidelity are significantly superior to other learned methods and approach those of G-PCC.

The qualitative comparison validates that learned point cloud attribute compression methods suffer from severe multi-generation distortion, while the proposed methods can effectively suppress this multi-generation distortion, significantly enhancing the robustness of learned compression frameworks in iterative compression scenarios. LCC, in particular, can achieve multi-generation robustness comparable to G-PCC.

\subsubsection{Quantitative Comparisons}
\begin{figure*}[!t]
\centering
\includegraphics[width=7.16in]{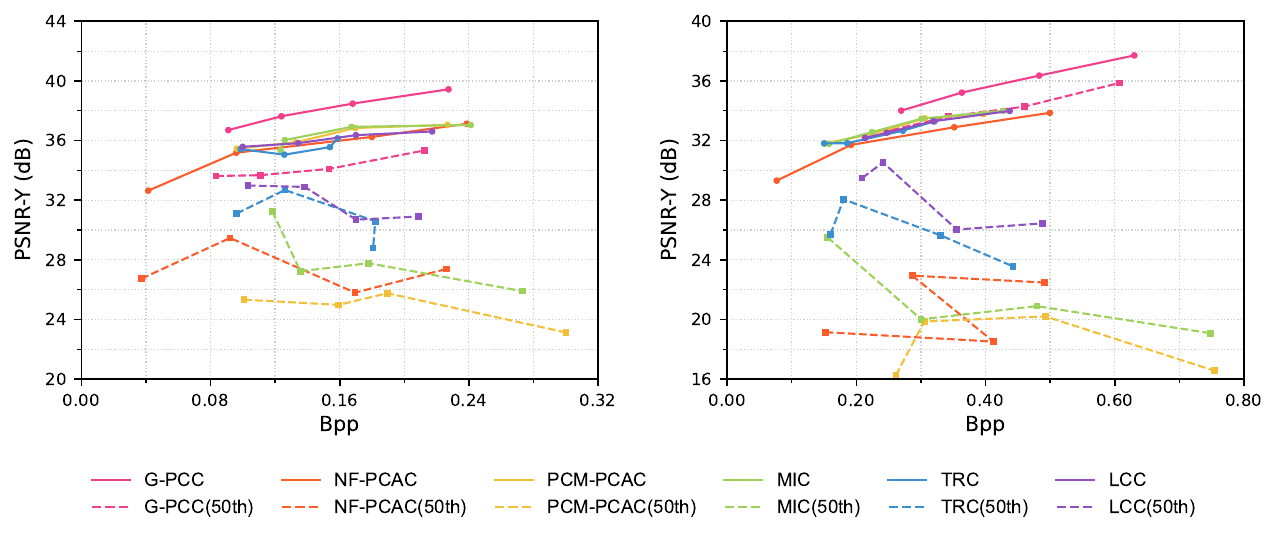}
\caption{Rate-distortion performance for single-pass (1st, solid lines) and 50-pass (50th, dashed lines) compression on the Owlii (left) and 8iVFB (right) datasets. Each color represents a different method.}
\label{fig_6}
\end{figure*}

In the previous subsection, we qualitatively evaluated the performance of various methods in multi-generation compression scenarios, validating the effectiveness of the proposed methods. In this section, we further verify the advantages of the proposed methods in multi-generation compression scenarios through more detailed quantitative comparisons. Fig. \ref{fig_6} shows the single-pass compression and multi-generation (50th) compression RD performance of the anchor methods and the proposed methods on the Owlii and 8iVFB datasets. Table~\ref{tab2} lists the specific quantitative evaluation results for multi-generation attribute compression for each method. By quantitatively comparing the RD performance, it can be seen that the proposed methods achieve RD performance closer to that of single-pass compression after 50 successive compression rounds compared to the baseline model. This significantly improves the reconstruction quality after multi-generation compression, thereby validating the effectiveness of the proposed methods. Furthermore, from Fig. \ref{fig_6}, it can also be observed that learned methods face more severe multi-generation distortion at high bitrates, including a decrease in reconstruction quality (PSNR-Y) and an increase in bitrate.

Among the proposed methods, MIC provides significant improvements in multi-generation compression robustness at low bitrates. Conversely, at high bitrates, its multi-generation compression performance degrades, and the gains over the baseline method are limited. We believe this because, under high bitrate conditions, the model retains more details of the original signal, especially high-frequency information. These high-frequency details are inherently more sensitive to minor perturbations during the transformation process. At this point, distortion accumulation dominated by transformation irreversibility may become the primary factor affecting multi-generation robustness. MIC primarily addresses the post-processing quantization perturbation problem in Section~\ref{section3B} by optimizing the complete end-to-end mapping, as defined by (\ref{Eq7}). Its contribution to suppressing cumulative distortion caused by transformation irreversibility is limited, making it difficult to improve multi-generation robustness at high bitrates.

TRC and LCC can significantly improve multi-generation compression robustness across all bitrates. TRC enhances the reversibility of the analysis/synthesis transformation by introducing a quantization-free auxiliary path, as defined by (\ref{Eq10}), alleviating the transformation irreversibility problem discussed in Section~\ref{section3C}. LCC suppresses the distortion accumulation and irreversible distortion of the latent space analyzed in Section~\ref{section3B} and Section~\ref{section3C} through an inter-generational compression auxiliary path, as defined by (\ref{Eq13}). These improvements not only enhance reconstruction quality in multi-generation scenarios but also ensure that the bitrate after multi-generation compression is similar to that of single-pass compression.

\begin{figure*}[!t]
\centering
\includegraphics[width=7.16in]{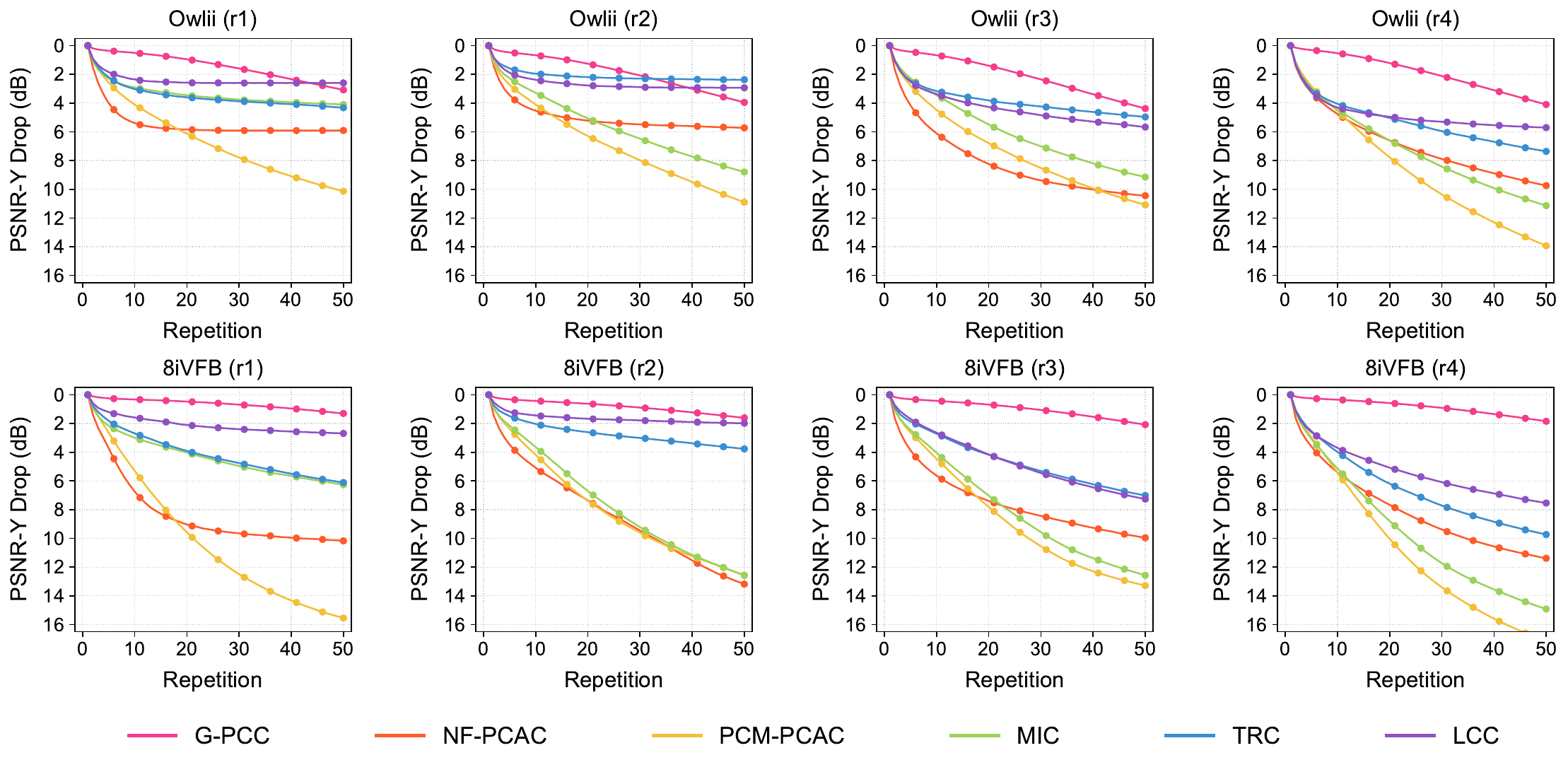}
\caption{Evolution of PSNR-Y Drop (dB) with increasing compression repetitions for various methods across different rate points (r1-r4) on Owlii (top) and 8iVFB (bottom) datasets.}
\label{fig_7}
\end{figure*}

\begin{table}[tb]
\centering
\captionsetup{skip=5pt} % Optional: add some space between caption and table
\caption{Quantitative Evaluation (PSNR-Y in dB) of Multi-generation Attribute Compression (1st and 50th passes)}
\label{tab2}
\renewcommand{\arraystretch}{1.2} % Increase row height slightly
\setlength\tabcolsep{5pt}
\begin{tabular}{@{}cl S[table-format=2.2] S[table-format=2.2] S[table-format=2.2] S[table-format=2.2] S[table-format=2.2] S[table-format=2.2]@{}}
\toprule
\multirow{2}{*}{Bpp} & \multirow{2}{*}{Method} & \multicolumn{2}{c}{Owlii} & \multicolumn{2}{c}{8iVFB} & \multicolumn{2}{c}{Average} \\
\cmidrule(lr){3-4} \cmidrule(lr){5-6} \cmidrule(lr){7-8}
                     &                         & {1st} & {50th} & {1st} & {50th} & {1st} & {50th} \\
\midrule
\multirow{6}{*}{r1}  & G-PCC                   & 36.71 & 33.62 & 34.01 & 32.71 & 35.36 & 33.17 \\
                     & NF-PCAC                 & 32.64 & 26.74 & 29.31 & 19.14 & 30.98 & 22.94 \\
                     & PCM-PCAC                & 35.46 & 25.33 & 31.80 & 16.26 & 33.63 & 20.79 \\
                     & MIC                     & 35.37 & 31.26 & 31.77 & 25.51 & 33.57 & 28.39 \\
                     & TRC                     & 35.42 & 31.10 & 31.81 & 25.70 & 33.62 & 28.40 \\
                     & LCC                     & 35.58 & 32.99 & 32.18 & 29.49 & 33.88 & 31.24 \\
\midrule
\multirow{6}{*}{r2}  & G-PCC                   & 37.63 & 33.68 & 35.22 & 33.61 & 36.43 & 33.65 \\
                     & NF-PCAC                 & 35.19 & 29.47 & 31.71 & 18.52 & 33.45 & 23.99 \\
                     & PCM-PCAC                & 35.88 & 24.98 & 32.43 & 19.86 & 34.16 & 22.42 \\
                     & MIC                     & 36.03 & 27.24 & 32.57 & 20.01 & 34.30 & 23.62 \\
                     & TRC                     & 35.06 & 32.68 & 31.83 & 28.06 & 33.44 & 30.37 \\
                     & LCC                     & 35.82 & 32.89 & 32.52 & 30.54 & 34.17 & 31.71 \\
\midrule
\multirow{6}{*}{r3}  & G-PCC                   & 38.48 & 34.10 & 36.36 & 34.28 & 37.42 & 34.19 \\
                     & NF-PCAC                 & 36.25 & 25.80 & 32.89 & 22.94 & 34.57 & 24.37 \\
                     & PCM-PCAC                & 36.84 & 25.76 & 33.49 & 20.20 & 35.17 & 22.98 \\
                     & MIC                     & 36.92 & 27.77 & 33.47 & 20.89 & 35.19 & 24.33 \\
                     & TRC                     & 35.56 & 30.59 & 32.66 & 25.65 & 34.11 & 28.12 \\
                     & LCC                     & 36.36 & 30.69 & 33.30 & 26.03 & 34.83 & 28.36 \\
\midrule
\multirow{6}{*}{r4}  & G-PCC                   & 39.44 & 35.34 & 37.71 & 35.87 & 38.58 & 35.60 \\
                     & NF-PCAC                 & 37.13 & 27.38 & 33.86 & 22.48 & 35.50 & 24.93 \\
                     & PCM-PCAC                & 37.06 & 23.13 & 33.77 & 16.56 & 35.42 & 19.84 \\
                     & MIC                     & 37.04 & 25.90 & 33.99 & 19.08 & 35.52 & 22.49 \\
                     & TRC                     & 36.15 & 28.79 & 33.29 & 23.56 & 34.72 & 26.17 \\
                     & LCC                     & 36.61 & 30.91 & 33.99 & 26.45 & 35.30 & 28.68 \\
\bottomrule
\end{tabular}
\end{table}
To focus more specifically on the multi-generation compression performance of each method, we use PSNR-Y Drop as the evaluation criterion, thereby eliminating interference from differences in initial reconstruction quality. Fig. \ref{fig_7} shows the Y-PSNR Drop performance of each method during multi-generation (50th) compression. It can be seen that LCC and TRC significantly improve multi-generation robustness, with LCC exhibiting better multi-generation compression performance than G-PCC in some cases. MIC can alleviate multi-generation distortion accumulation at low bitrates, but its performance is poorly at high bitrates, showing only a slight improvement over the baseline model. 

Furthermore, Fig. \ref{fig_7} also reveals that the multi-generation compression performance of learning-based methods is worse on low-resolution datasets (8iVFB) than on high-resolution datasets (Owlii), while the traditional compression method G-PCC shows the opposite trend. We speculate that this phenomenon may stem from the following points: Firstly, learned methods, especially models relying on deep convolutional networks, can more effectively learn and utilize rich texture details and spatial correlations when processing high-resolution data. When the resolution decreases, these models may struggle to fully leverage their learning capabilities due to insufficient input information, leading to reduced robustness against minor perturbations. Secondly, traditional methods like G-PCC typically employ block/region based transform coding. This mechanism might be less sensitive to global resolution, and its fixed transform bases and quantization strategies can maintain relatively stable performance when processing low-resolution data with comparatively simpler structures.

\begin{figure}[!t]
\centering
\includegraphics[width=3.5in]{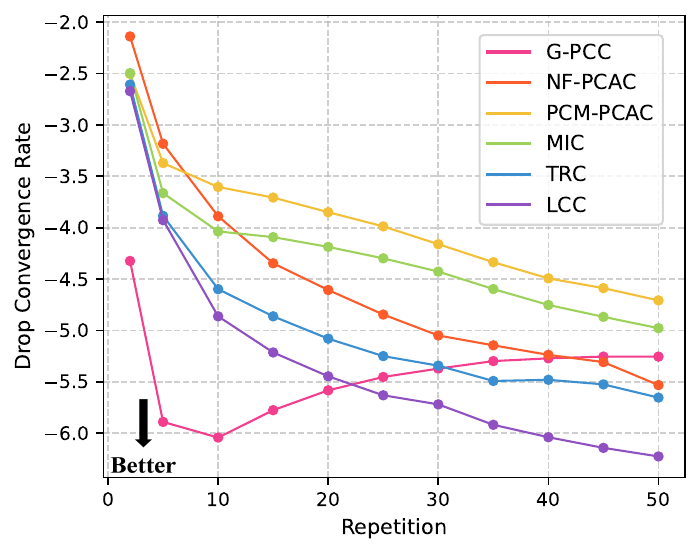}
% \caption{Average Drop Convergence Rate of Different Methods (averaged across all test sequences and rate points). A smaller Drop Convergence Rate indicates better convergence performance.}
\caption{Average Drop Convergence Rate for different methods, averaged across all test sequences and rate points. Smaller values indicate better convergence performance.}
\label{fig_8}
\end{figure}

Table~\ref{tab3} lists the specific average $\Delta$PSNR-Y results for each method after undergoing $k = \{2, 5, 10, 25, 50\}$ successive compression passes. To better illustrate the convergence speed of each method, we calculate the Drop Convergence Rate for each method using (\ref{Eq16}). After calculation, we selected the PSNR-Y Drop of PCM-PCAC after 50-pass successive compression as the Max PSNR-Y Drop. The average Drop Convergence Rate for each method is shown in Fig.~\ref{fig_8}. It can be observed that although G-PCC exhibits good multi-generation robustness in scenarios with a small number of repetitions, its $\Delta$PSNR-Y shows an upward trend after 10 successive compression rounds, making it difficult for the multi-generation compression quality to converge. In contrast, learned compression methods demonstrate a trend towards convergence in multi-generation compression quality, especially the proposed LCC method, which can achieve rapid convergence of $\Delta$PSNR-Y within a limited number of rounds. This rapid convergence characteristic indicates that the latent variable consistency constraint imposed by LCC can quickly guide the multi-generation compression process into a relatively stable state, avoiding the successive distortion accumulation discussed in Section~\ref{section3B}, and ultimately maintaining reconstruction quality close to that of G-PCC after 50-pass successive compression.

% \begin{table}
% \begin{center}
% \caption{Average $\Delta$ PSNR-Y (dB) Between Successive Generations at Different Repetition (k=2, 5, 10, 25, 50).}
% \label{tab3}
% \begin{tabular}{|c|c|c|c|c|c|}
% \hline
% \multirow{2}{*}{Method} & \multicolumn{5}{c|}{Repetition} \\
% \cline{2-6}
%                 & 2nd & 5th & 10th & 25th & 50th \\
% \hline
% G-PCC           & 0.1733       & 0.0362       & 0.0311        & 0.0561        & 0.0683        \\
% \hline
% NF-PCAC         & 1.5406       & 0.5422       & 0.2678        & 0.1029        & 0.0518        \\
% \hline
% PCM-PCAC        & 1.0696       & 0.4493       & 0.3561        & 0.2426        & 0.1181        \\
% \hline
% MIC             & 1.0781       & 0.3353       & 0.231         & 0.1778        & 0.0901        \\
% \hline
% TRC             & 0.9644       & 0.2688       & 0.1318        & 0.0687        & 0.0459        \\
% \hline
% LCC             & 0.9047       & 0.2579       & 0.1011        & 0.0469        & 0.0259        \\
% \hline
% \end{tabular}
% \end{center}
% \end{table}

\begin{table}[tb] % 您可以选择合适的浮动位置参数，如 [htbp], [!htb] 等
\centering
\captionsetup{skip=5pt} % 可选: 增加标题和表格之间的间距 (需要 caption 宏包)
\caption{Average $\Delta$PSNR-Y (dB) Comparison Between Different Successive Compression  Repetitions ($k = \{2, 5, 10, 25, 50\}$)} % 请替换为您的表格标题
\label{tab3}
\renewcommand{\arraystretch}{1.2} % 可选: 略微增加行高，使表格更疏朗
% 使用 S 列类型来按小数点对齐数字，table-format 定义了数字的格式
% l 代表第一列左对齐 (Method)
% @{} 用于去除列两边的额外空白
\begin{tabular}{@{}l S[table-format=1.4]
                   S[table-format=1.4]
                   S[table-format=1.4]
                   S[table-format=1.4]
                   S[table-format=1.4]@{}}
\toprule
\multirow{2}{*}[-0.3ex]{Method} & \multicolumn{5}{c}{ $\Delta$PSNR-Y in $k$-th Repetition ($\downarrow$)} \\
\cmidrule(lr){2-6} % (lr) 表示左右两边稍微缩短一些，更好看
                        & {2nd} & {5th} & {10th} & {25th} & {50th} \\ % S列的表头文本需要用{}括起来
\midrule
G-PCC                   & 0.1733 & 0.0362 & 0.0311 & 0.0561 & 0.0683 \\
NF-PCAC                 & 1.5406 & 0.5422 & 0.2678 & 0.1029 & 0.0518 \\
PCM-PCAC                & 1.0696 & 0.4493 & 0.3561 & 0.2426 & 0.1181 \\
MIC                     & 1.0781 & 0.3353 & 0.2310 & 0.1778 & 0.0901 \\ 
TRC                     & 0.9644 & 0.2688 & 0.1318 & 0.0687 & 0.0459 \\
LCC                     & 0.9047 & 0.2579 & 0.1011 & 0.0469 & 0.0259 \\
\bottomrule
\end{tabular}
\end{table}

Overall, the proposed methods can effectively improve the multi-generation compression robustness of learned point cloud attribute compression method. Among them, LCC exhibits the best multi-generation compression robustness, followed by TRC, while MIC is effective at low bitrates but performs poorly at high bitrates.

\subsection{Ablation Study}

\begin{figure}[!t]
\centering
\includegraphics[width=3.5in]{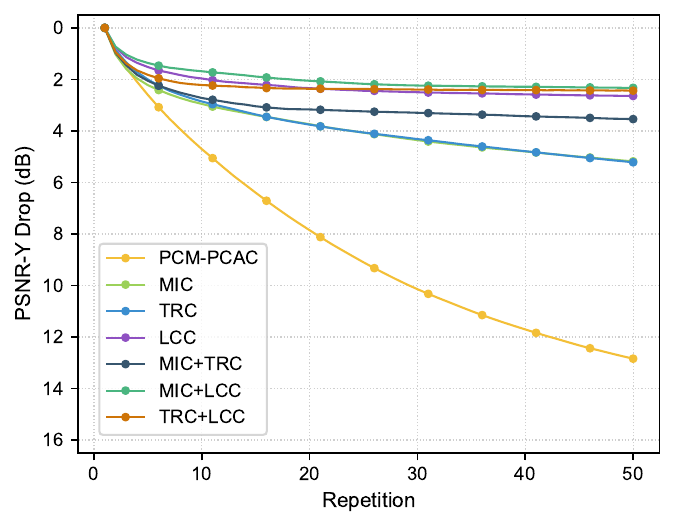}
\caption{Ablation study: average PSNR-Y Drop (dB) for combinations of proposed constraints on Owlii and 8iVFB datasets.}
\label{fig_9}
\end{figure}

In our ablation studies, we investigate the independence of the three proposed methods and the feasibility of their combinations. 

From an intuitive perspective of the loss functions, MIC replaces the original distortion loss $\mathcal{L}_D$ with $\mathcal{L}_{\text{MI}}$, while TRC and LCC are implemented by adding $\mathcal{L}_{\text{TR}}$ and $\mathcal{L}_{\text{LC}}$ respectively to the original loss function. We consider the following three pairwise combinations: MIC+TRC, MIC+LCC, and TRC+LCC. Combinations involving MIC replace the original distortion loss $\mathcal{L}_D$ with $\mathcal{L}_{\text{MI}}$, whereas the TRC+LCC combination adds both $\mathcal{L}_{\text{TR}}$ and $\mathcal{L}_{\text{LC}}$ to the original loss function. The loss functions corresponding to these combinations are shown as follow:
\begin{equation}
\label{Eq17}
\text{MIC+TRC: } \mathcal{L} = \mathcal{L}_{Rate} + \lambda \mathcal{L}_{\text{MI}} + \alpha \mathcal{L}_{\text{TR}},
\end{equation}
\begin{equation}
\label{Eq18}
\text{MIC+LCC: } \mathcal{L} = \mathcal{L}_{Rate} + \lambda \mathcal{L}_{\text{MI}} + \beta \mathcal{L}_{\text{LC}},
\end{equation}
\begin{equation}
\label{Eq19}
\text{TRC+LCC: } \mathcal{L} = \mathcal{L}_{Rate} + \lambda \mathcal{L}_D + \alpha \mathcal{L}_{\text{TR}} + \beta \mathcal{L}_{\text{LC}}.
\end{equation}

Fig. \ref{fig_9} shows the average PSNR-Y drop performance of the proposed methods and their three combinations during multi-generation compression on the Owlii and 8iVFB datasets at a bitrate point corresponding to $\lambda=1000$. It can be observed that the combined methods achieve better multi-generation compression performance than using any single method alone. The combination of MIC+LCC achieves the best multi-generation robustness improvement. At the same time, each combination method can produce additional gains, demonstrating that the proposed methods can independently contribute to alleviation, thereby validating their independence and effectiveness.

\section{Conclusion}\label{conclusion}
% In this paper, we, for the first time, 
This paper, for the first time, systematically investigate the multi-generation compression problem in point cloud attribute compression, revealing the core mechanisms by which learned point cloud attribute compression exhibits non-idempotency in multi-generation scenarios and the problems this causes. Based on this research analysis, we propose three novel and plug-and-play methods to enhance the robustness of multi-generation compression. Among them:

1) The Mapping Idempotency Constraint suppresses non-idempotency caused by incomplete training mapping by incorporating a post-processing step into the training path and adding a mapping idempotency loss term.

2) The Transformation Reversibility Constraint adds a quantization-free training path and a transformation reversibility loss term to the network, increasing the reversibility of the overall transformation in the network. 

3) The Latent Variable Consistency Constraint introduces a cross-generation compression path and a latent variable consistency loss term into the network, which suppresses cross-generation distortion accumulation while improving latent space consistency. 

Extensive experimental results on standard datasets clearly demonstrate that the proposed methods can effectively suppress distortion accumulation and maintain high-quality reconstruction details of point clouds in multi-generation compression scenarios, with their performance significantly outperforming existing baseline learning methods. In future work, we will extend the analytical framework and improvement ideas of this study to the equally critical problem of multi-generation compression for point cloud geometry, aiming to effectively enhance the practicality of point cloud compression in multi-generation scenarios.

% \section*{Acknowledgments}
% This should be a simple paragraph before the References to thank those individuals and institutions who have supported your work on this article.

% {\appendix[Proof of the Zonklar Equations]
% Use $\backslash${\tt{appendix}} if you have a single appendix:
% Do not use $\backslash${\tt{section}} anymore after $\backslash${\tt{appendix}}, only $\backslash${\tt{section*}}.
% If you have multiple appendixes use $\backslash${\tt{appendices}} then use $\backslash${\tt{section}} to start each appendix.
% You must declare a $\backslash${\tt{section}} before using any $\backslash${\tt{subsection}} or using $\backslash${\tt{label}} ($\backslash${\tt{appendices}} by itself
%  starts a section numbered zero.)}


\bibliographystyle{IEEEtran}
\bibliography{reference}

% \newpage

 

\begin{IEEEbiography}[{\includegraphics[width=1in,height=1.25in,clip,keepaspectratio]{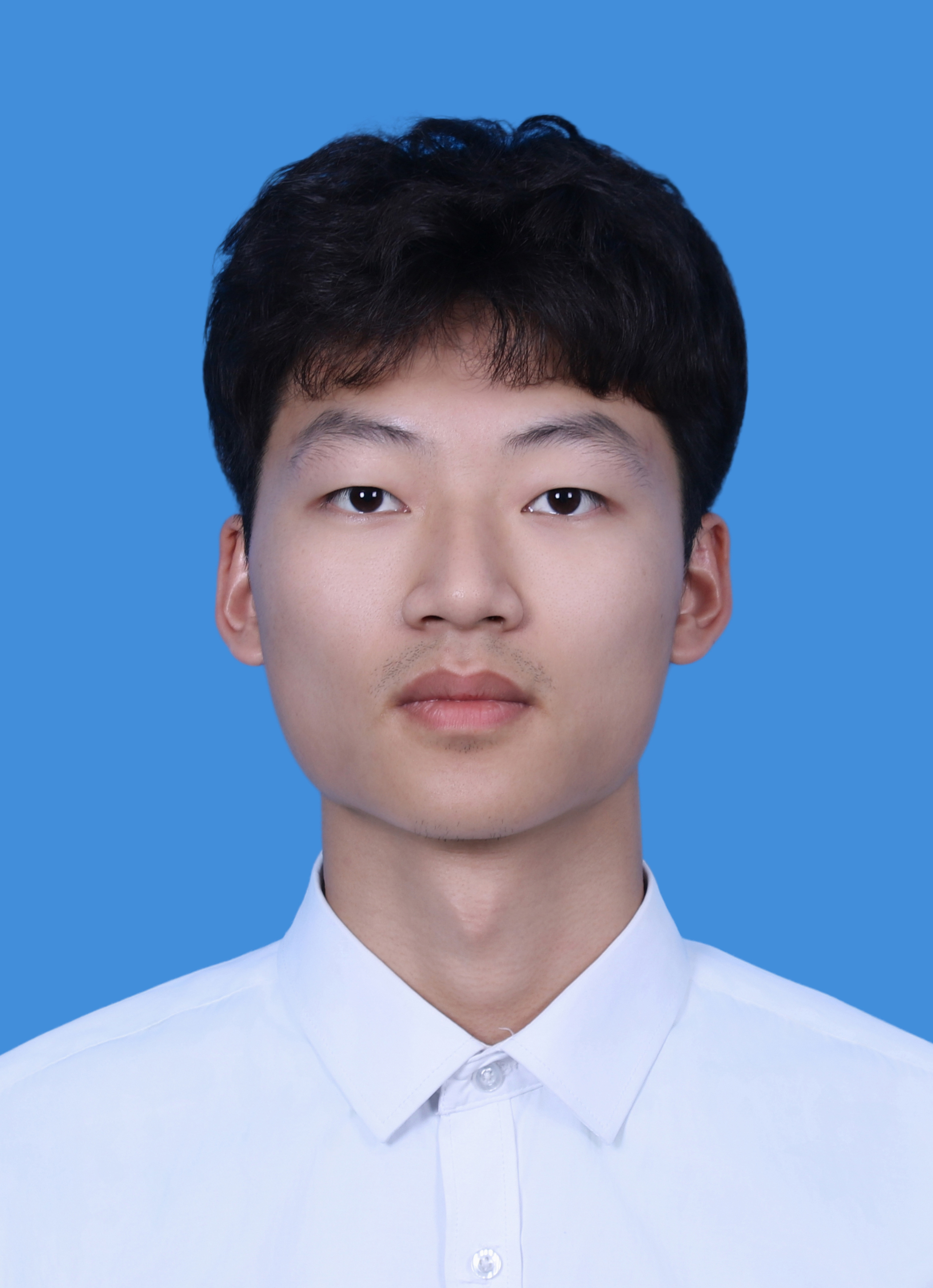}}]{Xiangzuo Liu}
received the B.Eng. degree from the School of Electronics and Information Technology, Sun Yat-sen University, Guangzhou, China, in 2024, where he is currently pursuing the M.Eng. degree in Information and Communication Engineering. His research interests include point cloud compression, video compression, and deep learning.
\end{IEEEbiography}

\vspace{11pt}

\begin{IEEEbiography}[{\includegraphics[width=1in,height=1.25in,clip,keepaspectratio]{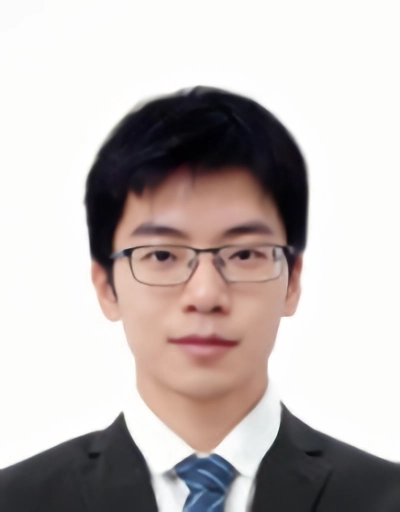}}]{Zhikai Liu} received the B.Eng. and M.Eng. degrees from Sun Yat-sen University, China, in 2022 and 2025, respectively. He is currently working in video coding standardization. His research interests include multimedia coding, processing and transmission.

\end{IEEEbiography}

\vspace{11pt}

\begin{IEEEbiography}[{\includegraphics[width=1in,height=1.25in,clip,keepaspectratio]{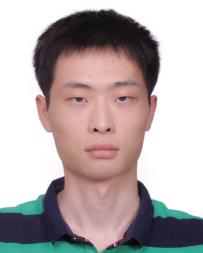}}]{Pengpeng Yu}
received the M.S. degree in Information and Communication Engineering from Sun Yat-sen University, Guangzhou, China, in 2023. He is currently working toward the Ph.D. degree in Electronic and Information Engineering with the School of Electronics and Communication Engineering, Sun Yat-sen University. His research interests include point cloud compression and computer vision.
\end{IEEEbiography}

\vspace{11pt}

\begin{IEEEbiography}[{\includegraphics[width=1in,height=1.25in,clip,keepaspectratio]{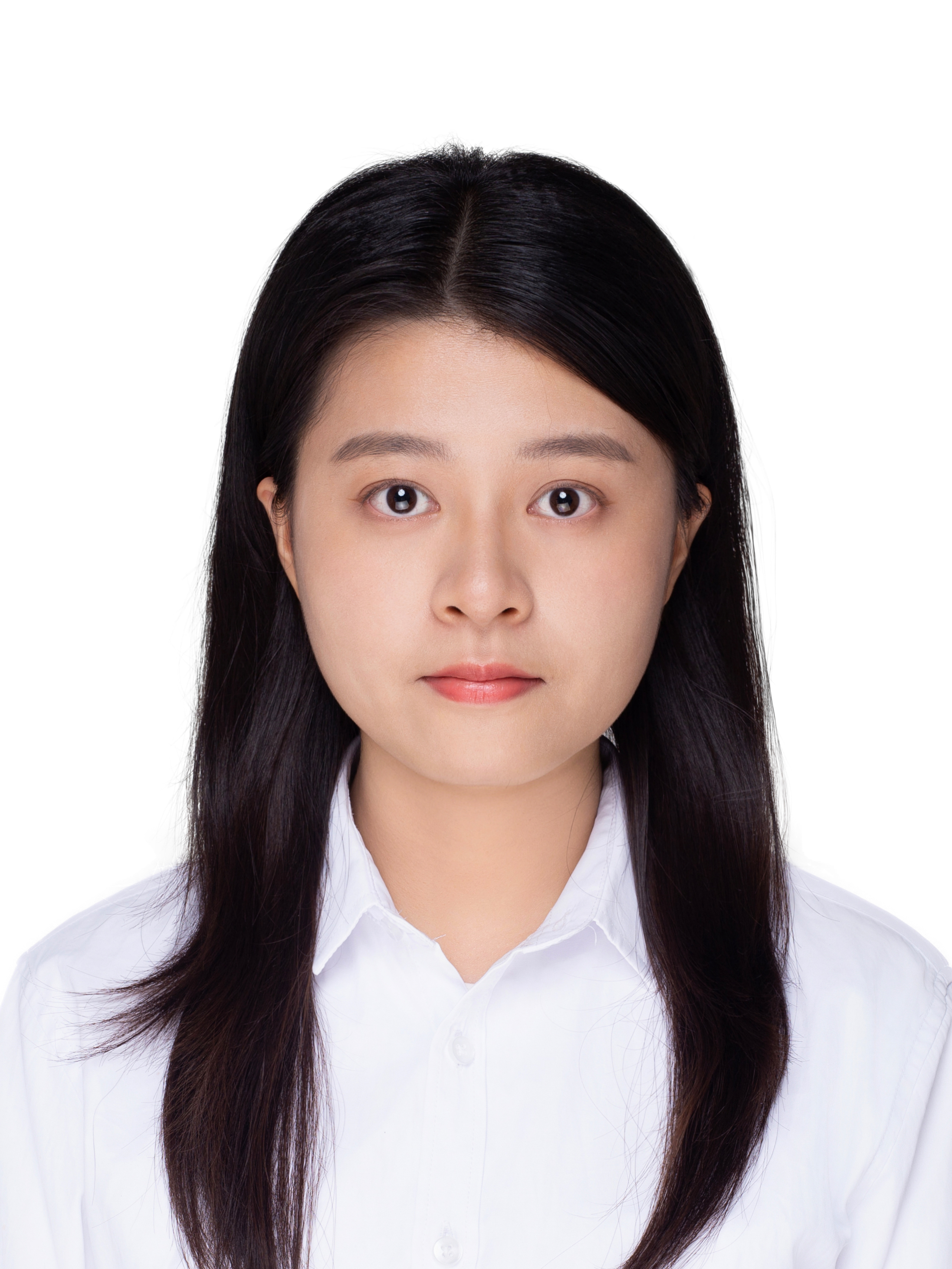}}]{Ruishan Huang}
received the M.Eng degree from the School of Electronics and Information Technology, Sun Yat-sen University, Guangzhou, China, in 2024. She is currently working in video compression. Her research interests include point cloud compression, video compression, and deep learning.
\end{IEEEbiography}

\vspace{11pt}

\begin{IEEEbiography}[{\includegraphics[width=1in,height=1.25in,clip,keepaspectratio]{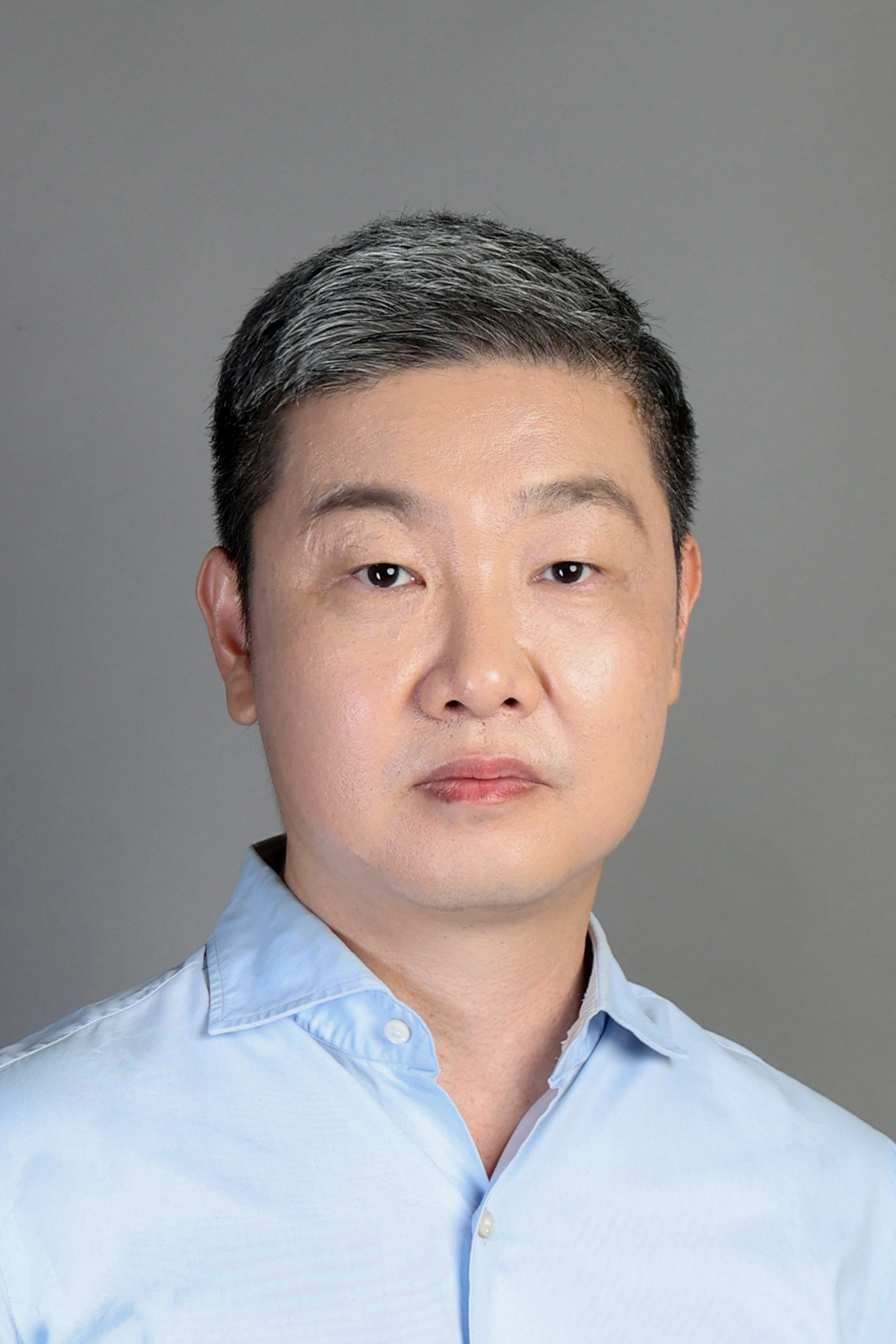}}]{Fan Liang}
(M’01) received the Ph. D. degree from Sun Yat-sen University, Guangzhou, China in 2000. He is currently an Associate Professor with Sun Yat-sen University. His current research interests include video coding, point cloud coding, multimedia communication, and atrificial intelligence techniques.
He is the Vice-Secretary General and the Chair of Requirements Subgroup of Audio Video Coding Standard Workgroup of China.
\end{IEEEbiography}


\vfill

\end{document}